\documentclass[prb,showpacs]{revtex4}
\usepackage{amsfonts,amsmath,epsfig}
\newcommand{\be}{\begin{equation}}
\newcommand{\ee}{\end{equation}}
\newcommand{\go}{\left<}
\newcommand{\gc}{\right>}
\newcommand{\po}{\left(}
\newcommand{\pc}{\right)}
\newcommand{\vo}{\left|}
\newcommand{\vc}{\right|}
\newcommand{\bo}{\left[}
\newcommand{\bc}{\right]}

\begin{document}
\title{Collective effects of the Coulomb interaction in anharmonic
quantum dots}
\author{D.~K.~Sunko}
\email{dks@phy.hr}
\affiliation{Department of Physics, Faculty of Science, University of
Zagreb,\\ Bijeni\v cka cesta 32, HR-10000 Zagreb, Croatia.}

\begin{abstract}

Deviations from the uniform oscillator spacing, related to the shape of the
confining potential, have a strong influence on few-electron states in quantum
dots when Coulomb effects are included. Distinct signatures are found for
level spacings increasing (pot shape) and decreasing (cone shape) with energy.
Cone deformations affect the levels near the ground state, such that
observable effects are predicted. Pot deformations partially negate the effect
of the Coulomb force, thus their spectra are similar to those of a perfect
oscillator with a smaller Coulomb repulsion. The Coulomb force is treated by
exact diagonalization, for which purpose efficient closed-form expressions for
the matrix elements are derived. The Coulomb matrix element in relative
coordinates is reduced to a single sum over four binomial coefficients, for
which simple analytic approximations are found.

\end{abstract}


\maketitle
\section{Introduction}

Quantum dots are both the smallest electronic devices, and the largest atoms.
Technologically, they allow the storage and manipulation of single electrons.
Physically, they are a practical laboratory to observe few-body Coulomb states
at energy scales far below those of real light atoms. In small quantum dots,
the Coulomb effects are typically about one-half of the oscillator spacing, so
they have been treated perturbatively with success,\cite{Warburton98} even
though they would appear a priori to belong to a crossover regime.

More recent developments have stressed the role of electronic correlations in
few-electron dots, among others with a view to manipulate quantum states for 
purposes of computation. Such efforts fall into two categories. One aims to
construct ``quantum dot molecules'' in particular geometric
patterns.\cite{Scheibner07} The other has focussed on ``electronic
molecules,'' referring to the limit of strongly interacting electrons within a
single quantum dot, where a molecule-like partition of energy into rotational
and vibrational degrees of freedom is observed.\cite{Kalliakos09} These latter
states might better be called ``quantum dot nuclei,'' since collectivization
by the repulsive Coulomb force occurs in the absence of centers of opposite
charge, which exist in molecules.

The present paper investigates the particular effect of well shape (in
vertical, not horizontal profile), modelled as departure from the uniform
single-particle energy spacing, on the collectivity of few-electron quantum
states within a single dot. The modelling is implemented simply by inserting
diagonal energy values, other than the ``correct'' oscillator ones. The two
qualitatively different cases are level spacings increasing and decreasing
with energy, referred to respectively as pot-shaped and cone-shaped dots.

The main result is that cone anharmonicities make quantum dots strongly prone
to creation of collective states. The physical reason is first demonstrated on
the classic case of two electrons (``quantum dot helium''), at modest values
of the Coulomb force, consistent with the literature.\cite{Merkt91} With four
electrons, there appear both a change in ordering of the lowest excited
states, and ``intruder'' states of exceptionally large spin and orbital
momentum among the higher ones. Both of these should lead to observable
effects, if such dots can be manufactured.

Two formal developments are also presented, which are not strictly necessary
for the proof of principle given here, but may be useful in larger
calculations. First, an advanced combinatorial technique\cite{Wilf92,Paule95}
is used to bring the center-of-mass (CM) Coulomb matrix elements to a more
tractable form. It gives rise to two simple analytical approximations, one of
them already known,\cite{Laughlin83-1} which give good estimates of the
Coulomb matrix elements over a large range of quantum numbers. Second, a
simple expression for the Coulomb matrix elements in laboratory
coordinates is obtained, based on an old symmetry-based
insight.\cite{Bakri67-1} The two together mean that binomial coefficients
suffice in Coulomb calculations for the two-dimensional oscillator: no
gamma or hypergeometric functions need apply.

\section{Center-of-mass matrix elements}

The 2D oscillator Hamiltonian is
\be
H_{\mathrm{osc}}=\left[-\frac{1}{2}\po 
\frac{\partial^2}{\partial r^2}+\frac{1}{r}\frac{\partial}{\partial r}
+\frac{1}{r^2}\frac{\partial^2}{\partial\phi^2}
\pc+
\frac{1}{2}r^2\right]\hbar\omega,
\label{hosc}
\ee
where the dimensionless coordinate $r$ has been scaled in units of the
oscillator characteristic length $l_0=\sqrt{\hbar/(m\omega)}$. It
has the eigenfunctions $\Psi_{nl}$,
\be
\Psi_{nl}(\mathbf{r})=\Psi_{nl}(x,y)=
\sqrt{\frac{n!}{\pi (n+l)!}}
L^{l}_{n}(x^2+y^2)(x+iy)^le^{-(x^2+y^2)/2},
\label{psiosc}
\ee
where $L^{l}_{n}$ is the generalized Laguerre polynomial, and there are
$2n+l+1$ degenerate eigenfunctions for each energy level:
\be
H_{\mathrm{osc}}\Psi_{nl}
=(2n+l+1)\hbar\omega\Psi_{nl}.
\ee
One can exploit a symmetry of the Laguerre polynomials,
\[
(-1)^k\frac{k!}{x^k}L_k^{n-k}(x)=(-1)^n\frac{n!}{x^n}L_n^{k-n}(x),
\]
to account for negative angular momenta in a natural way:
\be
\Psi_{n,-l}=(-1)^l\Psi^*_{n-l,l}.
\label{negang}
\ee
This allows one not to use absolute value signs on $l$, which is quite common
in the literature, but would ``break'' all the main formulae in the present
work. The notational price paid is that e.g. $\Psi_{11}$ and
$\Psi_{1,-1}=-\Psi^*_{01}$ do not belong to the same energy level: the
negative-momentum state corresponding to $\Psi_{11}$ is
$\Psi_{2,-1}=-\Psi^*_{11}$.

To calculate two-body matrix elements, transform as usual to the CM frame,
\be
\mathbf{R}=\frac{\mathbf{r}_1+\mathbf{r}_2}{\sqrt{2}},\quad
\mathbf{r}=\frac{\mathbf{r}_1-\mathbf{r}_2}{\sqrt{2}}.
\label{cm}
\ee
Denoting the CM basis $|NL;nl)=\Psi_{NL}(\mathbf{R})\Psi_{nl}(\mathbf{r})$, the
Coulomb interaction turns out to be
\begin{multline}
\po NL;nl\vc\frac{1}{|\mathbf{r}_1-\mathbf{r}_2|}\vo N'L';n'l'\pc=
\delta_{NN'}\delta_{LL'}\delta_{ll'}\sqrt{\frac{\pi}{2}}
\frac{\sqrt{n!n'!(n+l)!(n'+l)!}}{(n+n'+l)!}\po \frac{1}{4}\pc^{n+n'+l}
\\
\times
\sum_{j=0}^{\min(n,n')}\binom{2n+2n'-4j}{n+n'-2j}\binom{n+n'-2j}{n-j}
\binom{4j+2l}{2j+l}\binom{2j+l}{j},
\label{coulrel}
\end{multline}
where Zeilberger's algorithm\cite{Wilf92,Paule95} was used to reduce the
customary double sum\cite{Pfannkuche93} to a single one, as explained in
Appendix~\ref{zeilder}. Note that the sum has few terms in practice, since $n$
is at most one-half the principal quantum number $2n+l$. Dropping the CM
quantum numbers $N,L$, which introduces a notation for the relative-coordinate
wave functions, one observes
\be
\po n0\vc\frac{1}{|\mathbf{r}_1-\mathbf{r}_2|}\vo 00\pc=
\po 0n\vc\frac{1}{|\mathbf{r}_1-\mathbf{r}_2|}\vo 0n\pc,
\ee
which is quite noteworthy: the amplitude to scatter from the $2n$-th level to
the ground state is the same as the energy shift felt by the state of maximal
angular momentum in the $n$-th level. Thus the Coulomb force has the potential
to mix states very strongly, and this is enhanced by the cone-shape
anharmonicity, as explained below.

\begin{figure}
\raisebox{5mm}{\psfig{file=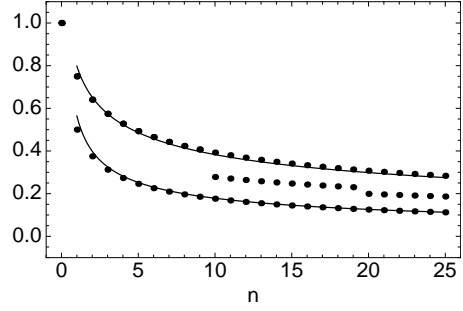,height=4cm}}
\caption{Coulomb matrix elements
$\po n0\vc\frac{1}{r\sqrt{2}}\vo n0\pc$ (upper points) and
$\po 0n\vc\frac{1}{r\sqrt{2}}\vo 0n\pc$ (lower points), along
with their analytic estimates (full lines). The intermediate case
$\po n,n/10\vc\frac{1}{r\sqrt{2}}\vo n,n/10\pc$ is also
shown for  $n\ge 10$ (the step at $n=20$ comes from discretizing $n/10$). All
matrix elements are in units of
$\left<00\vc\frac{1}{r\sqrt{2}}\vo 00\right>=\sqrt{\pi/2}$.
\label{figcoulomb}}
\end{figure}

To get some feeling for the values involved, note that the inequality
\be
\binom{2n}{n}<\frac{4^n}{\sqrt{n\pi}}
\label{binpi}
\ee
becomes an equality in the limit $n\to\infty$, i.e.\ is not an asymptotic
expression. This provides the approximation\cite{Laughlin83-1}
\be
\po 0n\vc\frac{1}{|\mathbf{r}_1-\mathbf{r}_2|}\vo 0n\pc
\sim\frac{1}{\sqrt{2n}}.
\ee
The energy shifts in states of smaller angular momentum are larger, but not by
much, the largest, derived in Appendix~\ref{approx}, being
\begin{equation}
\po n0\vc\frac{1}{|\mathbf{r}_1-\mathbf{r}_2|}\vo n0\pc\sim
\frac{1}{\sqrt{2n}}\left[
\sqrt{2}+\frac{\ln n}{\pi}
\right].
\end{equation}
As shown in Fig.~\ref{figcoulomb}, the range of values of the diagonal Coulomb
matrix elements is less than a factor of two, mostly given by the first
(square root) factor in brackets. The analytic estimates are remarkably good,
all the way down to $n=1$.

\section{Laboratory-frame matrix elements}

Many-body wave functions in laboratory coordinates best exploit the symmetry
between the particles. On the other hand, interaction terms in the Hamiltonian
are more easily calculated in the CM frame, as above. Denoting  the wave
function in laboratory coordinates $\vo
n_1l_1;n_2l_2\bc=\Psi_{n_1l_1}(\mathbf{r}_1) \Psi_{n_2l_2}(\mathbf{r}_2)$, we
need the transformation coefficients appearing in
\be
\vo NL;nl\pc=\sum_{n'=0}^{N+n}\quad\sum_{l'=-n'}^{N+L+n+l-n'}
C^{NL;nl}_{n'l'}\vo N+n-n',L+l-l';n'l'\bc.
\label{labcm}
\ee
They can be found following the observation\cite{Bakri67-1} that they must
also appear in the transformation of the simplest monomial having the required
symmetry, which is in this case
\[
r^{2n+l}e^{il\phi}=(x-iy)^n(x+iy)^{n+l}.
\]
One easily obtains, referring to Eq.~\eqref{cm},
\begin{multline}
(X-iY)^N(X+iY)^{N+L}(x-iy)^n(x+iy)^{n+l}=\\
\sum_{n'l'}d^{Nn}_{n'}d^{N+L,n+l}_{n'+l'}
(x_1-iy_1)^{N+n-n'}(x_1+iy_1)^{N+L+n+l-n'-l'}
(x_2-iy_2)^{n'}(x_2+iy_2)^{n'+l'}
\label{aux}
\end{multline}
where
\be
d^{Nn}_{n'}=\frac{1}{\sqrt{2^{N+n}}}\sum_{m=\max(0,n'-N)}^{\min(n,n')}(-1)^m
\binom{N}{n'-m}\binom{n}{m}.
\ee
It follows that
\be
C^{NL;nl}_{n'l'}=D^{Nn}_{n'}D^{N+L,n+l}_{n'+l'},\quad
D^{Nn}_{n'}=\sqrt{
\frac{\binom{N+n}{n}}
{\binom{N+n}{n'}}
}
d^{Nn}_{n'},
\ee
where the square-root factors are needed to remove the standard
normalizations of the wave functions, and Laguerre polynomials within them, so
that the highest power of $r$ appears with coefficient unity, for the $d$'s to
act upon as in Eq.~\eqref{aux}. The inverse of the CM transformation,
Eq.~\eqref{cm}, looks the same, so the same expression~\eqref{aux} must hold
upon exchanging $\mathbf{R}\leftrightarrow\mathbf{r}_1$ and
$\mathbf{r}\leftrightarrow\mathbf{r}_2$, meaning that ``)'' and ``]'' can
exchange places in Eq.~\eqref{labcm}. Thus, iterating Eq.~\eqref{labcm} is the
same as inverting it, i.e. the coefficients are idempotent:
\be
\sum_{N=0}^{n_1+n_2}\quad\sum_{L=-N}^{n_1+l_1+n_2+l_2-N}
C^{n_1l_1;n_2l_2}_{NL}
C^{n_1+n_2-N;l_1+l_2-L}_{N_2L_2}=\delta_{n_2N_2}\delta_{l_2L_2}.
\ee

Collecting the above results, one finds a computationally efficient
expression for the matrix element in laboratory coordinates,
\begin{multline}
\bo n_1l_1;n_2l_2\vc\frac{1}{|\mathbf{r}_1-\mathbf{r}_2|}
\vo n_3l_3;n_4l_4\bc=\\
\delta_{l_1+l_2,l_3+l_4}
\sum_{N=0}^{n_1+n_2}\quad
\sum_{L=-N}^{n_1+l_1+n_2+l_2-N}
C^{n_1l_1;n_2l_2}_{NL}C^{n_3l_3;n_4l_4}_{N+\Delta n,L}
\po NL\vc
\frac{1}{|\mathbf{r}_1-\mathbf{r}_2|}
\vo N+\Delta n,L\pc,
\end{multline}
where it is assumed that $\Delta n=n_3+n_4-n_1-n_2\ge 0$, without loss of
generality.

\section{The model}

The model used for shape effects is a simple modification of the
single-particle spectrum. The diagonal oscillator energies are modified
to read
\be
E(n,l,\delta)=(2n+l+1)^{1+\delta}\hbar\omega,
\label{edelta}
\ee
where $\delta=0$ is of course the original oscillator spectrum, and $\delta=1$
would correspond to the infinite-well spectrum, where the level energies
increase quadratically with the principal quantum number. Thus positive values
of $\delta$ are referred to as ``pot-shaped'' dots. Negative values of $d$
indicate level spacings decreasing with energy, and this is called
``cone-shaped.'' The extreme cone shape is the hydrogen spectrum, $\delta=-3$
(with a negative prefactor). The idea here is to vary $\delta$ in the more
modest range $\delta=\pm 0.5$, with some reasonable effective Coulomb strength.

Concerning the latter, the operative parameter is
\be
\alpha=\frac{1}{\epsilon}\frac{e^2/l_0}{\hbar\omega},
\label{coulr}
\ee
where $\epsilon$ is the relative dielectric constant of the parent material,
and $l_0=\sqrt{\hbar/(m\omega)}$ is the oscillator length scale, so that
$\alpha$ measures the Coulomb repulsion at well size in units of the
oscillator spacing.

To summarize, the model considered in the present work is given by the
Hamiltonian matrix
\be
\go A\vc H\vo B\gc=
\go A| B\gc
\sum_{(n,l)\in A}\frac{E(n,l,\delta)}{\hbar\omega}
+\alpha\go A\vc V_C\vo B\gc,
\label{model}
\ee
where $A,B$ are Slater determinants in configuration space and $E(n,l,\delta)$
is the modified harmonic-oscillator spectrum, Eq.~\eqref{edelta}. Calculations
are performed in the subspace of minimal spin projection (zero for even number
of electrons), which is first broken down into blocks by diagonalizing the
square of the total spin operator, thus identifying states by spin. This
achieves the same economy as if one had generated a good spin
basis\cite{Tureci06} at the beginning.

\section{Two electrons}

\begin{figure}
\includegraphics[height=10cm,angle=-90]{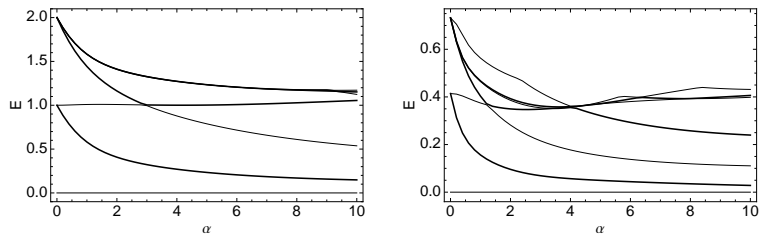}
\caption{Energy spectra of quantum dot helium, in dependence on the Coulomb
strength $\alpha$. Left is the recalculated result of
Ref.~\onlinecite{Merkt91}, with
$\delta=0$ in Eq.~\eqref{edelta}, and only the first six excited states shown,
the highest three degenerate. Right is the same with $\delta=-0.5$. All
energies are relative to the ground state, which appears as a horizontal line.
Successive levels in energy are drawn alternately by thin and thick lines.
Five oscillator shells are included. Note that each ``kink'' on the higher
curves in the right panel marks another level crossing, with even higher
states, not shown.
\label{figalpha}}
\end{figure}
Quantum dot helium, with only two electrons, is obviously not the best choice
to demonstrate collective effects. However, it already points to the reason
why cone deformations lead to enhanced collectivity. Figure~\ref{figalpha}
shows the evolution of the first six excited states as a function of Coulomb
strength, in the usually investigated range $\alpha<10$. 

There is an obvious overall scaling effect of the anharmonicity, so to compare
with the harmonic case, one should set energy windows to contain the same
number of states, rather than the same energy range. When this is done, as in
Fig.~\ref{figalpha}, a clear qualitative difference emerges. The cone
anharmonicity induces a much greater number of level crossings, and they shift
to lower values of the Coulomb force.

The reason for the effect is that the higher oscillator shells open a larger
phase space than the lower ones, and they fan out both up and down relative to
the ground state. Simply enough, the states going down cross sooner with those
coming up. Because there are many of them, the overall impact on the low-lying
states is quite substantial, resulting in a multitude of level crossings where
there have previously been only a few.

\section{Four electrons}

\begin{figure}
\includegraphics[height=15cm,angle=-90]{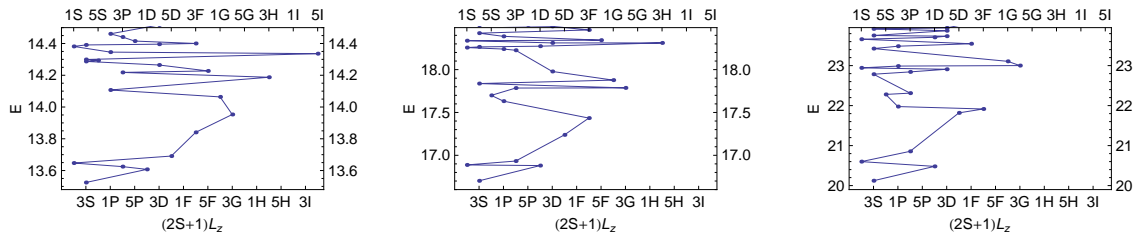}
\caption{Energy spectra of four electrons for $\alpha=3$ and three values of
$\delta$, left to right: $\delta=-0.5$, $\delta=0$, and $\delta=0.5$,
respectively. The ground state is Hund's rule triplet, denoted 3S. Note the
inversion in order of the next three states in the left panel, and a 5I
intruder, not appearing in the other two panels. The first $25$ states are
shown in all three panels, connected by a full line in order of increasing
energy. Four oscillator shells were used.
\label{figdelta}}
\end{figure}

Predictably, the effect described in the previous section becomes more
interesting with a larger number of electrons. Figure~\ref{figdelta} shows a
four-electron calculation, with $\delta=-0.5$, $\delta=0$, and $\delta=0.5$,
left to right. The first thing that meets the eye is the loss of high orbital
momentum states as $\delta$ increases, in particular the 3H state of the
oscillator disappears upwards for the pot anharmonicity. Conversely, a 5I
intruder moves down for the cone anharmonicity, to become the 16th excited
state, while it was 36th in the harmonic oscillator.

Closer inspection reveals two remarkable peculiarities of the cone. First,
while the 5I spin quintuplet has gone down in energy order, the 5S quintuplet,
also present in the two other panels, has gone up, such that it is nearly
degenerate with the 5I state of the cone. Calculation of electromagnetic
transitions is beyond the scope of this work, but it is clear from general
considerations that a 5S state lies in the decay path of a 5I state, because
one can turn into the other by shedding orbital momentum only, without
reconfiguring the spins. When they are nearly degenerate, decay of one into
the other is likely to be significantly supressed, because of the scarcity of
low-energy photons with the required high orbital momentum. This does not
prove, without a calculation, that the 5I will have an anomalously long
half-life, but does establish a plausible scenario for such an outcome.

The second cone peculiarity is the inversion of order of the first three
excited states, which is 1D--1S--3P for the harmonic oscillator, and
1D--3P--1S for the cone. The interest here is that these states have recently
been the subject of an extensive experimental analysis.~\cite{Kalliakos09}
Thus cone-anharmonic dots have a qualitative fingerprint, in principle
accesible to experiment, which may provide some motivation to try to
manufacture them, and distinguish them experimentally from harmonic or
pot-anharmonic ones.

The pot anharmonicity is much less interesting in comparison. Depletion of
higher oscillator shell contributions suppresses the effects of the Coulomb
force. The resulting spectra are similar to those obtained with a harmonic
oscillator with a weaker Coulomb interaction.

\section{Discussion and conclusion}

The main result of this work is that cone anharmonicities in the
single-particle level spacing of a quantum dot have a major impact on the
collective effects, when the Coulomb force is taken into account. In the
opposite limit, when the cone shape turns into a pot shape, the Coulomb force
can be partially negated, and the spectrum recovers a form similar to the
unperturbed oscillator with a smaller value of the interaction. In brief, cone
deformations enhance Coulomb collective effects, pot deformations suppress
them.

The physical mechanism leading to these effects is clear. Deviations from the
single-particle spectrum, considered in this work, affect the higher
oscillator shells more strongly. These are unoccupied in the non-interacting
limit, for the fillings considered here. The stronger the interaction, the
greater the admixture of the high shells to the ground state. Working against
it is the energy price of involving them to optimize the occupied low level
states. The benefit of even a slight reduction in their energy is high,
because the higher shells have large degeneracies, so they make a lot of
configuration space available for optimization. Conversely, when the component
of high shells decreases, the space for interactions becomes more restricted,
so the system behaves as if the Coulomb force were weaker than it is.

This interpretation is confirmed when one looks at the basis-vector
composition of the low cone states, such as the 5I state in
Fig.~\ref{figdelta}. Nothing dramatic happens with respect to the analogous
oscillator state, only the higher (in single particle ordering) components are
visibly more populated, and the lower ones, visibly less. However,
simultaneous energy and wave-function rearrangements can have an
order-of-magnitude impact on decay paths and lifetimes, providing a mechanism
to establish time-limited ``quantum protectorates''\cite{Laughlin00} with
tunable properties.

The calculations are schematic in the sense that harmonic-oscillator wave
functions do not diagonalize whichever potential would give the anharmonic
diagonal energies, introduced by hand into the model. Nevertheless, it is not
expected that the use of exact wave functions would affect the main outcome,
because the mechanism described above is robust to model details. Furthermore,
it is unlikely that the profile of a dot potential will ever be known with
precision.  The most that can be hoped is that some manufacturing prescription
will turn out anharmonic dots with strongly collective charge states. If this
happens, naturally one will assume that the potential well is cone-deformed.
One might argue that the converse has already happened, and that the success
of perturbation theory in describing real dots is partially due to the
tendency of current manufacturing procedures towards pot-like anharmonicities,
which suppress the collective effects of the Coulomb force.

While the effect itself is robust enough, its detailed outcomes appear to be
quite uncontrolled. However, in practice it would not be necessary to achieve
highly reproducible manufacturing to observe the effects discussed above.
(Quantum dot molecular arrays are currently also produced with a considerable
range of properties.\cite{Scheibner08}) All that would be required to first
order is that an acceptable percentage of manufactured dots have a tendency to
``get stuck'' in their de-excitation pathway. In any case, it is far-fetched
at present to speculate on the possible ramifications of the simple proof of
principle, given here. It is hoped, on the contrary, that its simplicity
and universality will motivate someone to try to obtain strongly collective
few-electron states by manufacturing anharmonic quantum dots with cone-like
confining potentials.

\appendix

\section{Coulomb interaction in relative coordinates\label{zeilder}}

The usual expression\cite{Pfannkuche93} for the Coulomb matrix element may be
rewritten
\be
\po nl\vc\frac{1}{r\sqrt{2}}\vo n'l'\pc
=\delta_{ll'}\sqrt{\frac{\pi}{2}}\sqrt{n!n'!(n+l)!(n+l')!}
\sum_{j=0}^{\min(n,n')}
\frac{1}{4^{2j+l}}\frac{(n+n'-2j)!}{(n-j)!(n'-j)!j!(l+j)!}S_j,
\ee
where
\be
S_j=\sum_{k=0}^{n+n'-2j}
\frac{(2l+2k+4j)!}{\bo(l+k+2j)!\bc^2(n+n'-2j-k)!}\frac{(-1)^k}{4^kk!}.
\ee
Zeilberger's algorithm\cite{Wilf92} gives\cite{Paule95}
\be
S_j=\frac{1}{16}\frac{(4j-2n-2n'+1)(4j-2n-2n'+3)}
{(4j+2l+1)(4j+2l+3)(2j-n-n')(2j-n-n'+1)}S_{j+1},
\ee
which could in principle be checked directly. The solution is immediate,
\be
S_j=\frac{(2n+2n'-4j)!(4j+2l)!}
{(n+n'-2j)!^2(n+n'+l)!(2j+l)!}\po\frac{-1}{4}\pc^{n+n'-2j}.
\ee
Inserting it above and rearranging terms into binomial coefficients gives
Eq.~\eqref{coulrel}. In checking the latter against the literature, one should
take into account the caution below Eq.~\eqref{negang}.

\section{Approximation for the zero-angular momentum matrix element
\label{approx}}

Using the approximation Eq.~\eqref{binpi} for the case $l=0$ in
Eq.~\eqref{coulrel}, one easily finds
\begin{align}
\po n0\vc\frac{1}{r\sqrt{2}}\vo n0\pc&\approx
\frac{\sqrt{n}}{2\pi\sqrt{2}}
\sum_{j=1}^{n-1}\frac{1}{j(n-j)}+\frac{\sqrt{2\pi}}{4^{2n}}\binom{4n}{2n}
\nonumber\\
&\approx
\frac{\sqrt{n}}{2\pi\sqrt{2}}\po
\sum_{j=1}^{n-1}\frac{1}{j(n-j)}+\frac{2\pi\sqrt{2}}{n}
\pc,
\end{align}
where the end terms in the sum are treated specially to preclude a nonsensical
result. The sum is approximated by an integral,
\[
\int_1^{n-1}\frac{1}{x(n-x)}dx=\frac{2\ln(n-1)}{n},
\]
and finally $\ln(n-1)$ is replaced by $\ln n$ by inspection, because it gives
a much better approximation.

\acknowledgments

Conversations with S.~Bari\v{s}i\'c are gratefully acknowledged. This work is
supported by the Croatian Government under Project~{119-1191458-0512}.

\end{document}